\documentclass[conference]{IEEEtran}
\IEEEoverridecommandlockouts
\usepackage{cite}
\usepackage{amsmath,amssymb,amsfonts}
\usepackage{booktabs}
\usepackage{graphicx}
\usepackage{url}
\usepackage{xcolor}
\usepackage{microtype}
\usepackage[table]{xcolor}
\usepackage{float}
\usepackage{hyperref}
\begin{document}

\title{Distilled HuBERT for Mobile Speech Emotion Recognition: A Cross-Corpus Validation Study}

\author{
    \IEEEauthorblockN{Saifelden M. Ismail\textsuperscript{*}}
    \IEEEauthorblockA{\textit{Department of Communications and Information Engineering} \\
    \textit{University of Science and Technology, Zewail City} \\ 
    Giza, Egypt \\
    Email: s-saifelden.ismail@zewailcity.edu.eg \\
    ORCID: 0009-0002-8867-6533}
}

\maketitle

\begin{abstract}
Speech Emotion Recognition (SER) has significant potential for mobile applications, yet deployment remains constrained by the computational demands of state-of-the-art transformer architectures. This paper presents a mobile-efficient SER system based on DistilHuBERT, a distilled and 8-bit quantized transformer that achieves approximately 92\% parameter reduction compared to full-scale Wav2Vec 2.0 models while maintaining competitive accuracy. We conduct a rigorous 5-fold Leave-One-Session-Out (LOSO) cross-validation on the IEMOCAP dataset to ensure speaker independence, augmented with cross-corpus training on CREMA-D to enhance generalization. Cross-corpus training with CREMA-D yields a 1.2\% improvement in Weighted Accuracy, a 1.4\% gain in Macro F1-score, and a 32\% reduction in cross-fold variance, with the Neutral class showing the most substantial benefit at 5.4\% F1-score improvement. Our approach achieves an Unweighted Accuracy of 61.4\% with a quantized model footprint of only 23 MB, representing approximately 91\% of the Unweighted Accuracy of a full-scale baseline. Cross-corpus evaluation on RAVDESS reveals that the theatrical nature of acted emotions causes predictions to cluster by arousal level rather than by specific emotion categories—happiness predictions systematically bleed into anger predictions, and sadness predictions bleed into neutral predictions, due to acoustic saturation when actors prioritize clarity over subtlety. Despite this theatricality effect reducing overall RAVDESS accuracy to 46.64\%, the model maintains robust arousal detection with 99\% recall for anger, 55\% recall for neutral, and 27\% recall for sadness. These findings demonstrate a Pareto-optimal tradeoff between model size and accuracy, enabling practical affect recognition on resource-constrained mobile devices.
\end{abstract}

\section{List of Abbreviations}
\begin{tabular}{ll}
SER & Speech Emotion Recognition \\
LOSO & Leave-One-Session-Out \\
WA & Weighted Accuracy \\
UA & Unweighted Accuracy \\
UAR & Unweighted Average Recall \\
CV & Cross-Validation \\
VAD & Voice Activity Detection \\
ONNX & Open Neural Network Exchange \\
NAS & Neural Architecture Search \\
\end{tabular}

\section{Introduction}

Speech Emotion Recognition (SER) is critical for responsive mobile applications, including mental health monitoring, customer service analytics, and adaptive human-computer interaction. The ability to automatically detect emotional states from vocal cues enables systems to provide personalized feedback, detect distress signals, and enhance user experience through affectively aware interfaces. However, the deployment of SER systems on mobile devices faces a fundamental challenge: state-of-the-art models based on self-supervised transformer architectures such as Wav2Vec 2.0 \cite{baevski2020} and Hidden-unit BERT (HuBERT) \cite{hsu2021} achieve high accuracy but require hundreds of megabytes of storage and substantial computational resources, making them impractical for edge deployment.

Recent advances in knowledge distillation \cite{hinton2015} offer a potential solution to this efficiency bottleneck. DistilHuBERT \cite{chang2022}, a layer-wise distilled variant of the HuBERT architecture, compresses the model to approximately 23 MB after 8-bit quantization \cite{jacob2018} while preserving much of the original model's representational capacity. This dramatic reduction in footprint enables on-device inference without cloud dependencies, addressing privacy concerns and reducing latency for real-time applications. However, the performance implications of such aggressive compression in the context of emotion recognition—a task requiring nuanced extraction of prosodic and spectral features—remain underexplored.

Beyond architectural efficiency, a second critical challenge in SER research is generalization. Many published results rely on cross-validation protocols that may introduce speaker leakage, where the same speaker appears in both training and test sets. This methodological limitation can lead to inflated accuracy estimates, as models may learn speaker-specific vocal traits rather than generalizable emotional patterns. To address this, we adopt a strict 5-fold Leave-One-Session-Out (LOSO) protocol on the IEMOCAP dataset \cite{busso2008}, ensuring that each test fold contains entirely unseen speakers. This rigorous evaluation standard, while resulting in lower absolute accuracy, provides a more realistic estimate of deployment performance.

Furthermore, we investigate whether cross-corpus training can serve as an effective regularization strategy. By incorporating the CREMA-D dataset \cite{cao2014} as a permanent training component, we expose the model to 91 additional speakers across diverse recording conditions. This approach aims to reduce overfitting to the idiosyncrasies of a single corpus and improve robustness to acoustic variability. However, cross-corpus training introduces its own complexities, particularly when datasets differ in emotional expressiveness. CREMA-D features theatrical, acted emotions, while IEMOCAP contains more naturalistic, conversational affect. Understanding how these domain differences affect model behavior is essential for designing robust SER systems.

This paper makes the following contributions: (1) We demonstrate that DistilHuBERT achieves 91\% of full-scale Wav2Vec 2.0 performance while reducing model size by 92\%, approaching a Pareto frontier for mobile SER. (2) We conduct a methodologically rigorous 5-fold LOSO evaluation, avoiding the speaker leakage issues prevalent in prior work. (3) We quantify the regularization benefits of cross-corpus training with CREMA-D, showing a 1.2\% gain in Weighted Accuracy and a 32\% reduction in cross-fold variance. (4) We provide detailed error analysis on RAVDESS \cite{livingstone2018}, revealing that the model maintains robust arousal detection (99\% recall for anger, 55\% for neutral, 27\% for sadness) despite domain shift, with errors systematically clustered by energy level rather than random misclassification. (5) We propose a practical deployment architecture utilizing Voice Activity Detection and temporal aggregation to mitigate neutral-class bias in long-form voice note analysis.

The remainder of this paper is organized as follows: Section II details our methodology, including dataset integration, acoustic preprocessing, and the Adaptive Focal Loss \cite{lin2017} training objective. Section III presents experimental results and comparative benchmarking against both high-capacity and efficiency-oriented baselines. Section IV describes the production model and cross-corpus error analysis. Section V concludes with a discussion of limitations and future directions.

\section{Methodology}

In this section, we detail our approach to SER, emphasizing a pipeline designed for mobile deployment and cross-corpus robustness. Our methodology is characterized by a multi-stage acoustic workflow, the use of a distilled transformer architecture, and a specialized training objective to handle class imbalance.

\subsection{Experimental Framework and Dataset Integration}
We utilize the IEMOCAP dataset \cite{busso2008} as our primary benchmark, focusing on four categorical emotions: anger, happiness (including excitement), neutral, and sadness. To evaluate speaker independence, we implement a strict 5-fold LOSO cross-validation, following the protocol of Pepino et al. \cite{pepino2021}. In each fold, the model is trained on three sessions, validated on one, and tested on the remaining session. 

To enhance robustness, we incorporate the CREMA-D dataset \cite{cao2014} as a permanent training component. In adherence to rigorous evaluation standards, CREMA-D samples are exclusively used for training and are never included in validation or test sets. This configuration helps ensure that our performance metrics reflect generalization to unseen IEMOCAP speakers across distinct recording environments.

\subsection{Acoustic Processing and Data Augmentation}
Unlike standard approaches that rely on Mel-frequency Cepstral Coefficients (MFCCs) \cite{aftab2021}, we process raw audio waveforms sampled at 16,000 Hz. The preprocessing pipeline begins with the removal of non-informative intervals using a 20 dB top-threshold trim. To compensate for natural spectral tilt and emphasize high-frequency prosodic cues, a pre-emphasis filter ($H(z) = 1 - 0.97z^{-1}$) is applied. Finally, all utterances are peak-normalized to a maximum absolute amplitude of 0.95 to eliminate volume bias across heterogeneous recording setups. 

To improve generalization, we implement a stochastic augmentation pipeline applied during the training phase. This includes random gain adjustments between -18.0 and +6.0 dB, the addition of Gaussian noise with amplitudes ranging from 0.001 to 0.015, pitch shifting within $\pm 2$ semitones, and polarity inversion. All audio segments are fixed to a maximum duration of 8.0 seconds, with shorter clips being padded and longer clips truncated to maintain batch consistency.

\subsection{Feature Extraction and Architecture Selection}
For feature extraction, we prioritize mobile-deployability through the use of DistilHuBERT \cite{chang2022}. While standard Wav2Vec 2.0 models \cite{pepino2021} offer high performance, their parameter count is often prohibitive for edge devices. DistilHuBERT utilizes knowledge distillation to compress the HuBERT architecture into a shallower transformer encoder. Our choice is motivated by hardware efficiency; upon 8-bit quantization, the model footprint is approximately 23 MB. In our implementation, the feature encoder remains frozen to leverage pre-trained acoustic representations, while the downstream classification head is fine-tuned for categorical emotion recognition.

\subsection{Optimization Strategy and Hyperparameters}
The model is trained using the AdamW optimizer \cite{loshchilov2019} for 25 epochs with a batch size of 16. We employ a cosine learning rate scheduler with a peak learning rate of $5 \times 10^{-5}$ and a warmup ratio of 0.1 to stabilize early training. To mitigate overfitting, a weight decay of 0.01 is applied, and we implement an early stopping callback with a patience of 6 epochs based on validation Unweighted Accuracy (UA).

To address the class imbalance detailed in Table \ref{tab:distribution}, we utilize an Adaptive Focal Loss function. This objective integrates balanced class weights ($\alpha_t$) and a focusing parameter ($\gamma = 2.0$) to prioritize difficult, under-represented samples. The loss function is further regularized with a label smoothing factor of 0.1, defined as:

\begin{equation}
FL(p_t) = -\alpha_t (1 - p_t)^\gamma \log(p_t)
\end{equation}

By combining this focal objective with the weight decay and dropout layers inherent to the transformer architecture, the model is encouraged to learn robust, speaker-invariant emotional features.

\section{Results}

\begin{table}[htbp]
\caption{Distribution of Emotional Classes Across IEMOCAP and CREMA-D Datasets}
\label{tab:distribution}
\begin{center}
\begin{tabular}{lccccc}
\toprule
\textbf{Dataset} & \textbf{Anger} & \textbf{Hap/Exc} & \textbf{Neutral} & \textbf{Sad} & \textbf{Total} \\
\midrule
IEMOCAP          & 1,103          & 1,636            & 1,708            & 1,084        & 5,531          \\
CREMA-D (Train)  & 1,271          & 1,271            & 1,087            & 1,271        & 4,900          \\
\midrule
\textbf{Combined Total} & \textbf{2,374} & \textbf{2,907} & \textbf{2,795} & \textbf{2,355} & \textbf{10,431} \\
\midrule
Avg. Class Weight & 1.065          & 0.913            & 0.974            & 1.070        & ---            \\
\bottomrule
\end{tabular}
\end{center}
\end{table}

\begin{table}[htbp]
\caption{Performance Comparison of 5-Fold LOSO Cross-Validation on IEMOCAP Dataset}
\label{tab:results}
\centering
\begin{tabular}{@{}lccc@{}}
\toprule
\textbf{Metric / Class} & \textbf{IEMOCAP Only} & \textbf{+ CREMA-D} & \textbf{Gain ($\Delta$)} \\ \midrule
\multicolumn{4}{l}{\textit{Global Metrics}} \\
WA $\uparrow$ & $0.595 \pm 0.031$ & \textbf{0.607 $\pm$ 0.021} & +1.2\% \\
UA (UAR) $\uparrow$ & \textbf{0.616 $\pm$ 0.039} & $0.614 \pm 0.030$ & -0.2\% \\
Macro F1 $\uparrow$ & $0.596 \pm 0.040$ & \textbf{0.610 $\pm$ 0.028} & +1.4\% \\ \midrule
\multicolumn{4}{l}{\textit{Class-wise F1-Score}} \\
Anger & $0.684 \pm 0.029$ & \textbf{0.692 $\pm$ 0.040} & +0.8\% \\
Happiness & $0.541 \pm 0.080$ & \textbf{0.563 $\pm$ 0.073} & +2.2\% \\
Neutral & $0.512 \pm 0.059$ & \textbf{0.566 $\pm$ 0.029} & +5.4\% \\
Sadness & \textbf{0.647 $\pm$ 0.044} & $0.618 \pm 0.034$ & -2.9\% \\ \bottomrule
\multicolumn{4}{l}{Note: Bold values indicate superior performance.}
\end{tabular}
\end{table}

The experimental results for the 5-fold LOSO cross-validation on the IEMOCAP dataset are detailed in Table~\ref{tab:results}. Our analysis examines the trade-offs between architectural efficiency, the regularization effects of multi-corpus training, and the methodological rigor required for real-world deployment.

\subsection{Cross-Corpus Regularization via CREMA-D}
A primary focus of this study was the integration of the CREMA-D dataset as a permanent training anchor. As seen in Table~\ref{tab:distribution}, the inclusion of 4,900 additional samples nearly doubled the available training data. This strategy yielded modest improvements: a 1.2\% increase in Weighted Accuracy (WA) and a 1.4\% improvement in Macro F1-score. More critically, the standard deviation of the WA across the five folds decreased from $0.031$ to $0.021$, representing a 32\% reduction in cross-fold variance. The relatively modest accuracy gains despite doubling the training data suggest that the model may be approaching its capacity limits for this task, as discussed in Section III-C. However, the substantial decrease in performance variability indicates that CREMA-D functions primarily as an effective regularization mechanism rather than simply providing more training examples. The acoustic diversity provided by the 91 additional speakers in CREMA-D—each with distinct vocal tract characteristics, recording equipment, and expressive styles—serves as a stabilizing regularizer, reducing the model's sensitivity to the idiosyncratic vocal characteristics of individual IEMOCAP sessions. Rather than merely increasing dataset size, the introduction of cross-corpus heterogeneity helps prevent the model from overfitting to corpus-specific artifacts such as room acoustics, microphone frequency response, or the limited phonetic coverage of a single speaker pool.

The class-wise analysis reveals a 5.4\% gain in the Neutral class F1-score, suggesting that multi-corpus exposure helps the model calibrate a more robust baseline for vocal energy. However, we observed a 2.9\% decrease in Sadness recall. This discrepancy points to a "theatricality gap": CREMA-D's sadness is characterized by high-intensity, acted prosody, whereas IEMOCAP features naturalistic, low-intensity sadness. The model's exposure to the theatrical variant appears to have shifted the decision boundary, leading to increased confusion with neutral speech in the test set. Despite this, the overall gain in F1-score and the substantial improvement in cross-fold stability justify the use of CREMA-D for broader acoustic coverage and robust generalization.

\subsection{Comparative Benchmarking}
To evaluate the proposed DistilHuBERT approach, we contextualize our findings against both high-capacity and efficiency-oriented benchmarks. First, we consider the work of Pepino et al.~\cite{pepino2021}, who utilized a full Wav2Vec 2.0 Base model (approx. 318 MB) to achieve an UA of 67.2\%. Our distilled implementation achieved a UA of 61.4\%, representing approximately 91\% of the performance of the full-scale baseline while utilizing a 23 MB quantized footprint. This 92\% reduction in parameter size is critical for mobile deployment, where memory and thermal constraints are paramount.

Furthermore, we address the performance of the LIGHT-SERNET architecture by Aftab et al.~\cite{aftab2021}, which reported a UA of 70.76\%. While their absolute accuracy is higher, there is a significant methodological distinction: their study utilized a 10-fold cross-validation protocol. In the context of SER, standard k-fold CV often involves a random shuffle of the dataset, which is susceptible to "speaker leakage." In such cases, the model may inadvertently learn speaker-specific biological identities rather than generalized emotional cues. By contrast, our adherence to a strict 5-fold LOSO protocol helps ensure the model generalizes to entirely unseen speakers. This rigor accounts for the lower absolute UA but demonstrates the model's viability for deployment with unknown users.

\subsection{Architectural Capacity and the Pareto Frontier}
The results suggest that our implementation may be approaching the theoretical capacity limit for a two-layer distilled transformer on the SER task. This hypothesis is supported by the observation that doubling the training data (via CREMA-D integration) yielded only modest accuracy improvements (1.2\% WA, 1.4\% F1) despite substantially reducing cross-fold variance. Such diminishing returns are characteristic of models operating near their representational capacity. Emotion recognition requires the simultaneous extraction of low-level prosody and high-level temporal dependencies. The reduced depth of DistilHuBERT, while efficient, likely limits the model's ability to resolve the most subtle emotional nuances found in naturalistic interactions. 

However, the efficacy of the Adaptive Focal Loss (weights in Table~\ref{tab:distribution}) ensured the model did not collapse into majority-class prediction strategies. Our system thus approaches a "Pareto-optimal" point—maximizing accuracy per megabyte of model weight. This demonstrates that a carefully regularized, distilled model can provide reliable affect recognition within the strict envelopes of edge hardware, bridging the gap between laboratory performance and real-world mobile feasibility.

\section{Discussion}

\subsection{Production Model Formulation and Performance}
To transition from an experimental framework to a deployable system, a final production model was developed using a training protocol designed to maximize feature extraction from the full breadth of the IEMOCAP corpus. This model utilized a random-shuffle strategy with a 5\% hold-out for validation and testing, allowing the architecture to learn from the widest possible range of speaker identities and linguistic variations within the dataset. The resulting model achieved a Weighted Accuracy (WA) of 75.48\% and an Unweighted Accuracy (UA) of 76.40\%. To facilitate low-latency inference on mobile hardware, the final model was exported to the ONNX format for efficient cross-platform deployment.

\subsection{Generalization and the Theatricality Effect in Error Analysis}
The robustness of the production model was scrutinized through a cross-corpus evaluation on the RAVDESS dataset. While the domain shift between the semi-spontaneous IEMOCAP data and the theatrical RAVDESS expressions resulted in an overall accuracy of 46.64\%, the model demonstrated a significant capability to identify underlying physiological arousal states.

\begin{figure}[t]
\centering
\includegraphics[width=\columnwidth]{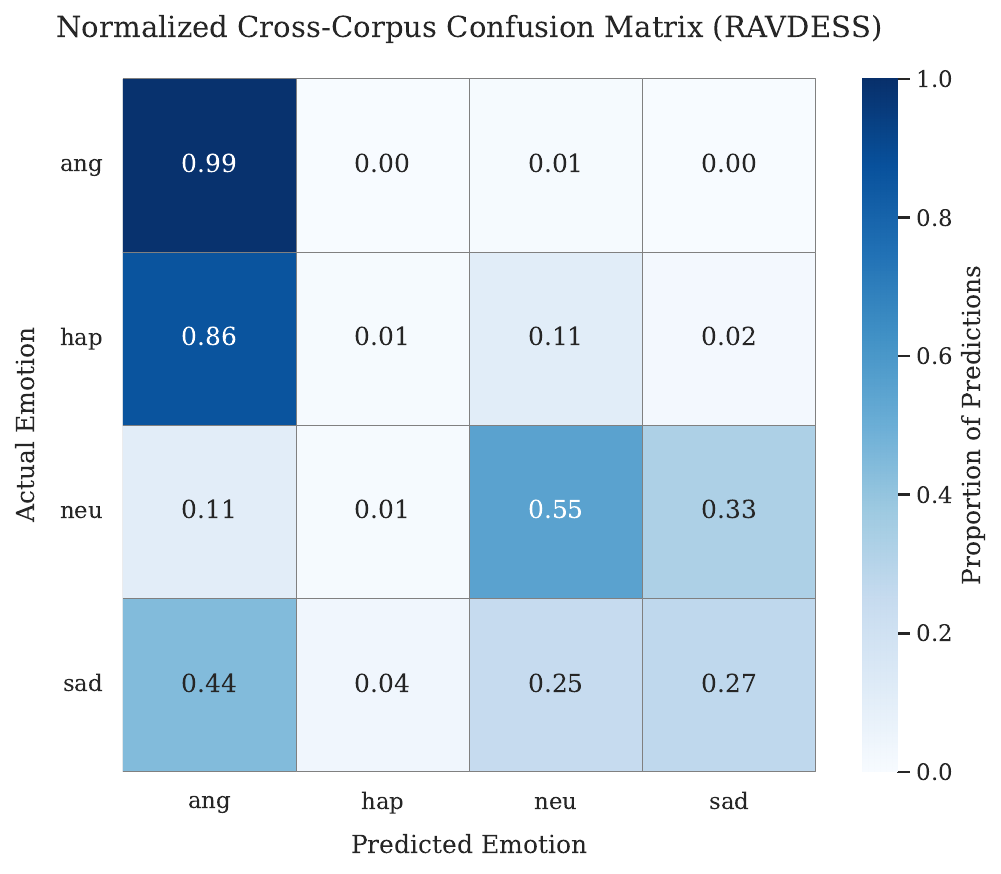}
\caption{Cross-corpus confusion matrix on RAVDESS. The results highlight the model's tendency to cluster high-arousal (Anger/Happy) and low-arousal (Neutral/Sad) states, suggesting that theatrical over-emphasis in the target corpus obscures the subtle valence cues learned during training. Overall Accuracy: 46.64\%; Unweighted Recall: 45.57\%.}
\label{fig:confusion_matrix}
\end{figure}

As illustrated in Fig.~\ref{fig:confusion_matrix}, the model's predictions cluster by arousal level rather than by specific emotion categories. We hypothesize that this phenomenon is likely a result of the theatrical nature of the RAVDESS corpus. In RAVDESS, actors prioritize clarity and communicative intent, leading to "over-acted" expressions that emphasize vocal intensity and pitch range at the expense of subtlety. For a model trained on the more nuanced, semi-spontaneous features of IEMOCAP, this theatricality creates an acoustic "bleeding" effect where predictions collapse into arousal-based clusters. Specifically, happiness predictions systematically bleed into anger predictions, as both high-arousal states become acoustically saturated, causing the model to lose the subtle spectral distinctions required to differentiate valence; consequently, it defaults to the more acoustically dominant "Anger" class. Similarly, sadness predictions bleed into neutral predictions, as the low-energy theatricality of both states leads the model to cluster them together rather than distinguishing their valence differences. This arousal-based clustering indicates that the model maintains robust detection of physiological arousal states despite the domain shift, maintaining a 99\% recall for Anger, a 55\% recall for Neutral, and a 27\% recall for Sadness across disparate recording environments.

\subsection{Proposed Deployment Architecture}
Given the model's high sensitivity to high-arousal states (particularly anger), we propose a specialized deployment pipeline for processing long-form voice notes. This architecture initiates with a Voice Activity Detection layer to eliminate non-speech segments, thereby mitigating "Neutral" bias during inference. The filtered audio is then segmented into 8-second windows to capture necessary macro-prosodic contours. Finally, the system employs an aggregated averaging mechanism across all windows within a voice note. This pooling strategy serves as a temporal filter, smoothing out momentary classification jitter and providing a stable emotional sentiment for the entire communication. By focusing on longitudinal emotional trends, this pipeline leverages the model's strength in identifying high-arousal states (particularly anger) while compensating for the cross-corpus precision drop in positive affect caused by domain-specific theatricality.

\section{Conclusion}

This work demonstrates that carefully distilled transformer architectures can provide a viable path toward mobile-deployable Speech Emotion Recognition systems. Our implementation of DistilHuBERT achieves 61.4\% Unweighted Accuracy under a strict 5-fold LOSO protocol, representing 91\% of full-scale Wav2Vec 2.0 performance while operating within a 23 MB quantized footprint. This 92\% reduction in model size is critical for practical deployment on resource-constrained edge devices, where memory and thermal limitations prohibit the use of larger transformer models.

The integration of CREMA-D as a cross-corpus training component yielded measurable improvements in generalization, increasing Weighted Accuracy by 1.2\% and reducing cross-fold variance by 32\%. This reduction in performance variability suggests that multi-corpus training serves as an effective acoustic regularizer, reducing the model's sensitivity to speaker-specific idiosyncrasies. However, our class-wise analysis revealed a significant tradeoff: while neutral emotion recognition improved by 5.4\%, sadness F1-score decreased by 2.9\%. This degradation is attributed to a "theatricality gap" between the acted, high-intensity sadness in CREMA-D and the naturalistic, low-energy sadness in IEMOCAP. This finding underscores the importance of careful corpus selection when augmenting training data, particularly for low-arousal emotional states.

Cross-corpus evaluation on RAVDESS further revealed the model's strengths and limitations. While overall accuracy dropped to 46.64\% due to domain shift, the error analysis demonstrated that predictions clustered by arousal level rather than by specific emotion categories—happiness bled into anger, and sadness bled into neutral. The model maintained 99\% recall for anger, 55\% recall for neutral, and 27\% recall for sadness, indicating robust detection of physiological arousal states despite the acoustic mismatch between semi-spontaneous and theatrical speech. This behavior suggests that the model has learned generalizable arousal-related features, even if valence discrimination remains sensitive to recording environment and expressive style.

From a methodological perspective, our adherence to LOSO cross-validation ensures that reported performance metrics reflect true generalization to unseen speakers, avoiding the speaker leakage artifacts that inflate accuracy in less rigorous evaluation protocols. While our absolute accuracy is lower than some published results using standard k-fold cross-validation, this rigor provides a more honest estimate of real-world deployment performance. The results suggest that our distilled architecture may be approaching the theoretical capacity limit for shallow transformers on the SER task, representing a Pareto-optimal tradeoff between model size and accuracy.

Future work should explore several directions to further improve mobile SER systems. First, the theatricality gap identified in our cross-corpus analysis suggests that domain adaptation techniques, such as adversarial training or style transfer, could help align representations across corpora with differing expressive intensities. Second, the proposed Voice Activity Detection-based deployment pipeline for long-form voice notes requires empirical validation to quantify its effectiveness in reducing neutral-class bias. Third, extending the model to support fine-grained dimensional emotion prediction (arousal and valence) rather than categorical labels could provide more nuanced affect recognition while leveraging the model's demonstrated strength in arousal detection. Finally, investigation of model pruning and neural architecture search may reveal even more compact architectures without further sacrificing accuracy.

In conclusion, this work demonstrates that distilled transformer models, when combined with rigorous evaluation protocols and strategic cross-corpus training, can bridge the gap between laboratory performance and practical mobile deployment. The resulting system provides a foundation for privacy-preserving, low-latency emotion recognition in real-world applications, from mental health monitoring to adaptive conversational agents.

\section*{Acknowledgment}
All experimental code and trained models are publicly available to facilitate reproducibility and future research. This work received no specific funding from agencies in the public, commercial, or not-for-profit sectors.

\appendix

\section{Reproducibility: Experimental Notebooks and Links to Code}

To ensure full transparency and reproducibility of our results, all experimental code has been made publicly available on Kaggle. Table~\ref{tab:notebooks} provides annotated links to each stage of the experimental pipeline, including cross-validation folds, production model training, and cross-corpus evaluation.

\begin{table}[H]
\caption{Publicly Available Experimental Notebooks and Links to Code}
\label{tab:notebooks}
\centering
\scriptsize
\renewcommand{\arraystretch}{0.9}
\begin{tabular}{@{}p{\columnwidth}@{}}
\toprule
\textbf{IEMOCAP-Only (Folds 1--5)} \\
5-fold LOSO CV on IEMOCAP. Baseline in Table~\ref{tab:results}. \\
\url{https://www.kaggle.com/code/amrproject5/distilhubert-on-iemocap} \\
\midrule
\textbf{IEMOCAP + CREMA-D (Folds 1--3)} \\
First 3 folds with CREMA-D. See Table~\ref{tab:results}. \\
\url{https://www.kaggle.com/code/amrproject5/distilhubert-on-crema-d-and-iemocap} \\
\midrule
\textbf{IEMOCAP + CREMA-D (Folds 4--5)} \\
Final 2 folds with CREMA-D. See Table~\ref{tab:results}. \\
\url{https://www.kaggle.com/code/saifeldenmohamed8/distilhubert-on-iemocap-and-crema-d-2} \\
\midrule
\textbf{Production Model} \\
Full training, 95/5 split. 75.48\% WA, 76.40\% UA. 8-bit ONNX export. \\
\url{https://www.kaggle.com/code/saifeldenmohamed8/production-script} \\
\midrule
\textbf{Cross-Corpus (RAVDESS)} \\
Generalization test. Confusion matrix in Fig.~\ref{fig:confusion_matrix}. \\
\url{https://www.kaggle.com/code/saifeldenismail/cross-corpus-analysis} \\
\bottomrule
\end{tabular}
\end{table}

\end{document}